\documentclass[preprint,superscriptaddress,preprintnumbers,amsmath,amssymb,floatfix,showpacs,prl]{revtex4}

\usepackage{graphicx}
\usepackage{dcolumn}
\usepackage{bm}
\usepackage{ulem} 
\usepackage{multirow} 

\begin{document}

\title{Marginally Unstable Periodic Orbits in Semiclassical Mushroom Billiards}
\author{Jonathan Andreasen}
\affiliation{
  Department of Applied Physics, Yale University,
  New Haven, CT 06520, USA
}
\author{Hui Cao}
\affiliation{
  Department of Applied Physics, Yale University,
  New Haven, CT 06520, USA
}
\author{Jan Wiersig}
\affiliation{
  Institute for Theoretical Physics, University of Magdeburg, 
  D-39016 Magdeburg, Germany 
} 
\author{Adilson E. Motter}
\affiliation{
  Department of Physics and Astronomy, Northwestern University,
  Evanston, IL 60208, USA
}

\date{\today}

\begin{abstract} 
Optical mushroom-shaped billiards offer a unique opportunity to isolate and
study semiclassical modes concentrated on non-dispersive, marginally unstable 
periodic orbits.  
Here we show that the openness of the cavity to external electromagnetic 
fields leads to unanticipated consequences for the far-field radiation 
pattern, including directional emission. 
This is mediated by interactions of marginally unstable periodic orbits 
with chaotic modes.
We also show that the semiclassical modes are robust against perturbations 
to the shape of the cavity, despite the lack of structural stability of 
the corresponding classical orbits. 
\end{abstract}
\pacs{05.45.-a, 05.45.Mt}

\maketitle

The study of chaotic billiards has served as a foundation to many areas, 
ranging from nonlinear dynamics and statistical physics to quantum optics. 
Classically, the phase space of a chaotic billiard, defined by its 
position and momentum variables, is typically divided into chaotic and 
regular regions, which is a property found in generic Hamiltonian systems.
However, in contrast with smooth dynamical systems, several billiards have 
been shown to have families of marginally unstable periodic orbits (MUPOs) 
embedded in their chaotic regions \cite{altmann08}. 
These orbits, of which bouncing-ball orbits \cite{vivaldi83} form a 
particular case, are zero-volume structures reminiscent of 
Kolmogorov-Arnold-Moser islands and dynamically indistinguishable from 
regular orbits \cite{altmann06}.  
Of the billiards exhibiting this property, no other is attracting as much 
attention as the mushroom-shaped billiards \cite{bunimovich01}.

Mushroom billiards have the distinctive non-generic feature of exhibiting a 
single chaotic and a single regular region \cite{bunimovich01}. 
These billiards have been used as a model to address the impact of MUPOs 
on chaotic orbits \cite{altmann08,altmann06,altmann05} and to examine
quantum tunneling \cite{backer08,dietz07,barnett07,vidmar07}.  
Yet, the fundamental role of MUPOs in semiclassical dynamics remains 
largely unexplored.  
The very physical reality of semiclassical MUPO modes remains to be 
demonstrated given that classical MUPOs are not structurally robust 
against parameter perturbations likely to be present in realistic 
situations \cite{altmann06,zapfe08}.

In this Letter, we investigate a family of MUPOs in optical dielectric 
mushroom cavities.  
Previous studies have used total internal reflection to identify lasing modes 
based on stable \cite{gmachl98,bermudez03} and unstable periodic orbits \cite{lee02} 
in chaotic cavities, where the latter correspond to the so-called {\it scarred} 
modes. 
In both cases the orbits of interest may not correspond to the least leaky 
modes of the passive system, but give rise to lasing modes via manipulation 
of the gain medium. 
In contrast, here we show that even in the absence of gain, total internal 
reflection can be used to trap selected MUPOs inside the billiard. 
All the other orbits leave the cavity through refraction, making the MUPO modes
the least leaky modes.
We explore computationally the consequences of this scenario in the 
semiclassical regime (i.e., short-wavelength regime).
We show that the reentry of electromagnetic fields into the cavity leads to a 
coupling between MUPOs and chaotic modes, which has no analogue in closed 
billiards and significantly impacts the directionality and intensity of the 
far-field radiation pattern. 
We also show that the semiclassical MUPO modes are robust against roughness and 
other perturbations to the geometry of the billiard. 

\begin{figure}
  \includegraphics[width=4cm]{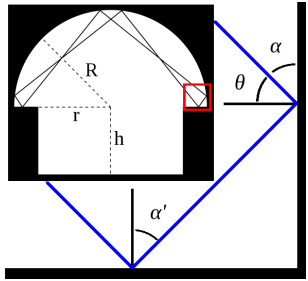} 
  \caption{\label{fig:fig1} (Color online)
    Bottom corner of the hat of the mushroom cavity.
    The ray represents the trajectory of an orbit with reflection
    angle $\theta=90^{\circ}-\alpha$ that hits the undercarriage with angle 
    $\alpha' \le \alpha$.
    When close enough to the corner, the curvature can be ignored and 
    $\alpha'=\alpha$.
    The inset shows a period-4 MUPO and the entire mushroom cavity with the 
    parameters $R$, $r$ and $h$ measured from the center of the diameter of 
    the semicircular hat.
    This point is also the origin of the cylindrical and Cartesian coordinates 
    used throughout this Letter, in which $z$ will represent the axis 
    perpendicular to the plane of the billiard.
  }
\end{figure}

We consider mushroom billiards composed of a circularly shaped hat with
radius $R$ and a foot, square or triangular in shape, that extends to a 
radial position of $r$ (Fig. \ref{fig:fig1}).
Regular orbits are confined to the hat between the circumferences of 
radii $r$ and $R$. 
MUPOs are also confined to the hat but necessarily cross the circle of radius 
$r$, while chaotic orbits necessarily visit the foot \cite{bunimovich01}. 
In the phase space (defined by the coordinates of arc length and reflection 
angle in the circumference of the hat), the regular and chaotic regions are 
separated by the reflection angle 
$\theta^* =  \sin^{-1}(r/R)$ \cite{bunimovich01,altmann05}.
The MUPOs have a reflection angle lying in the chaotic region 
($\theta < \theta^*$) but have null Lyapunov exponents, 
assuring a non-dispersive behavior for nearby trajectories.
An infinite number of families of MUPOs, characterized by different periods 
and rotation numbers, have been shown to exist for almost all choices of 
$r/R$ \cite{altmann08,altmann05}.
Here we focus on the particular family of period-$4$ MUPOs 
(Fig. \ref{fig:fig1}), which exist for $r/R>1/\sqrt{2}$ and have a 
reflection angle $\theta_p = 45^{\circ}$.

These MUPOs can be trapped and isolated in an open mushroom cavity with 
certain refractive index $n$ by exploiting the critical incidence 
angle $\theta_c = \sin^{-1}(1/n)$ below which light escapes the cavity.
Isolating the target MUPOs requires that all chaotic and regular modes leave
the cavity.
The former is easily accomplished since chaotic orbits will eventually have a 
reflection angle smaller than the critical angle, as in previous applications
to select stable modes in chaotic resonators \cite{gmachl98}.
The latter, on the other hand, can be accomplished by forcing the regular orbits to
escape through the ``undercarriage'' (bottom of the hat), as demonstrated next.

Figure \ref{fig:fig1} shows the trajectory of an orbit. 
If the reflection is close enough to the corner, the curvature of the hat can
be ignored and $\alpha' = \alpha$.
The MUPOs and the refractive index are chosen such that the reflection angle
$\theta_p$ satisfies $\theta_p > \theta_c$ and
$\alpha'_p = 90^{\circ} - \theta_p > \theta_c$.
We also want a cavity in which $\alpha'_{\mbox{\scriptsize{regular}}} < \theta_c$.
The smallest reflection angle for any regular orbit in the mushroom cavity is
$\theta^*$. 
Therefore, our conditions become
$90^{\circ} - \theta^* < \theta_c < 90^{\circ} - \theta_p$ and $\theta_c < \theta_p$.

We shall consider two commonly used materials in microcavity fabrication: 
polymer and semiconductor (Table \ref{table1}).
We first focus on polymer, which has refractive index $n=1.5$ 
and critical angle $\theta_c \approx 41.8^{\circ}$. 
We assure  $90^{\circ} - \theta^* < \theta_c $ by taking $r/R=0.75$, 
which yields $90^{\circ}-\theta^* \approx 41.4^{\circ}$.
Thus, since $\theta_p=45^{\circ}>\theta_c$, our conditions are met and 
the period-4 MUPOs will be isolated in the cavity.

\begin{table}[tbh]
  \caption{Parameters of the materials, period-4 MUPOs, cavities, 
    and FDTD grid spacing considered. 
    \medskip}
  \begin{tabular} {lccccccc}
    \hline
    & $n$ & $\theta_c$ & $\theta_p$ & $\theta^*$ &$~r/R~$ & $h/R$ & ~$\lambda/( n\Delta x )$\\
    \hline
    polymer & $1.5$ & $41.8^{\circ}$ & $45^{\circ}$ & $48.6^{\circ}$ & $0.75$ & $0.70$& $20$ \\ 
    \hline
    semiconductor & $3.3$ & $17.6^{\circ}$ & $45^{\circ}$ & $ 46.9^{\circ}$ & $0.73$ &$0.68$ &$20$\\ 
    \hline
    \label{table1}
  \end{tabular}
\end{table} 

The above argument assumes we are in the classical regime and is expected to be 
approximately valid in the semiclassical regime considered here, which is 
characterized by $nkR \gg 1$ and hence small wavelengths $\lambda=2\pi/k$ compared 
to the size $R$ of the cavity.
We set the parameters of the cavity to be $R=10\mu$m, $r=7.5\mu$m, and $h=7\mu$m. 
We consider TE polarization ($E_z = 0$) and employ a parallelized version of 
the finite-difference time-domain (FDTD) method \cite{tafl05} in 2D
with grid spacing of $\Delta x=\Delta y=1.36\,$nm for a square grid in
the plane of the cavity.
The cavity is excited uniformly with a sinusoidal-Gaussian pulse and the 
simulation run until only the eigenmodes with the highest 
$Q\equiv\omega/\gamma$ remain, where $\omega$ is the angular frequency and 
$\gamma$ is the energy decay rate of each eigenmode.

The eigenmodes are determined by a Fourier transformation of the energy in the 
cavity. 
A wide range of excitation wavelengths were simulated before arriving at the 
highest $Q$ profile for $\lambda=41.0$ nm.
The high $Q$ modes are limited to such small wavelengths most likely because
of the distance the MUPO modes are restricted to traveling in the corner of 
the mushroom hat. 

Because we work with the TE polarization, our far-field analysis deals with 
$H_z$. 
The field outside the cavity as a function of the angle $\phi$ at radial distance 
$\rho$ from the center of the diameter of the hat behaves as 
$H_z(\phi) = \sum_m a_m H_m(k\rho) e^{im\phi},$ where $H_m$ is the Hankel function 
of the first kind with azimuthal order $m$ \cite{jackson_classical_1998}.
The coefficients $a_m$ are found by applying 
$2\pi a_m H_m(k\rho)= \Delta\phi \sum_j H_z(\phi_j)e^{-im\phi_j},$ where 
$\phi_j$ is the angle discretized according to a resolution $\Delta\phi$.
The field is then evaluated at a far-field distance $\rho_f$, determined by the 
Fraunhofer condition $\rho_f \gg R^2/\lambda$, to obtain the far-field pattern 
$|H_z^f(\phi_j)|^2$.
In general $H_z(\phi_j)$ is complex, but FDTD only calculates the real component
of the field. 
We obtain the imaginary component from Faraday's law
$\mbox{Im} H_z = (-\omega\mu_0)^{-1} \nabla \times E$, where $E$ is real.

\begin{figure}[t]
  \includegraphics[width=8cm]{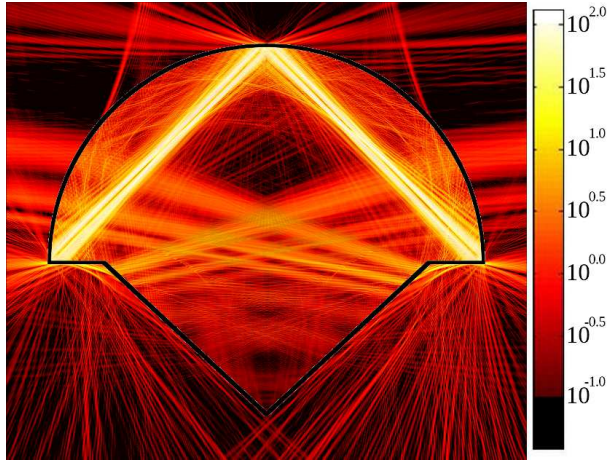} 
  \caption{\label{fig:fig2} (Color online)
    Spatial intensity $|H_z|^2$ for the period-$4$ MUPO mode with
    $\lambda = 41.0\,$nm and $Q = 13\, 000$ in a polymer microcavity.
    The intensity was averaged over one oscillation period.
  }
\end{figure}

Figure \ref{fig:fig2} color-codes the field intensity for a polymer cavity with a triangular
foot, indicating that the highest $Q$ mode is in this case a single period-4 MUPO mode 
with $\lambda=41.0\,$nm. 
A fit to the mono-exponential curve of energy versus time yields a decay rate $\gamma$ 
that  results in $Q=13\,000$. 
This eigenmode was isolated by narrowing the excitation width to $\delta\lambda = 0.005\,$nm. 
In closed billiards,  the shape of the foot does not play a crucial role. 
In the open cavities considered here, however, the shape of the foot
can significantly influence the dynamics of the modes. 
This is particularly so for polymer cavities since their modes are prone to leak.
For example, transitioning from a triangular foot to a square foot results 
in the MUPOs no longer being the highest $Q$ modes~\cite{comment}:
new high $Q$ modes are established via leakage from the undercarriage 
and subsequent penetration into the foot. 
This is expected to be relevant to optical experiments because the highest $Q$ mode 
typically determines the first lasing mode. 
More strikingly, when the MUPO remains the highest $Q$ mode, as in the  triangular-foot 
example of Fig.~\ref{fig:fig2},  field reentry can lead to {\it directional emission}.

\begin{figure} 
  \begin{tabular}{rl}
    \multirow{1}{*}[1.8cm]{ 
      \includegraphics[width=4cm]{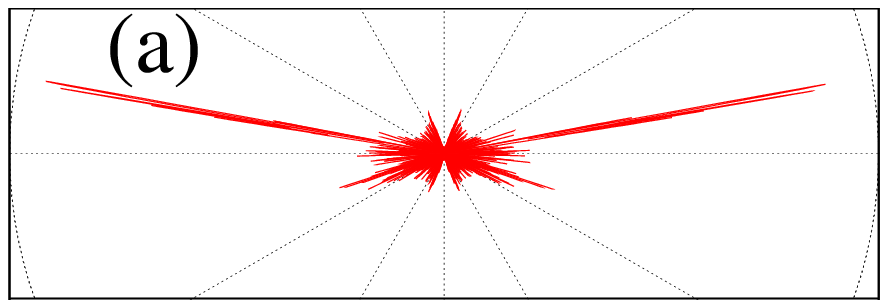} 
    } & \includegraphics[width=4cm]{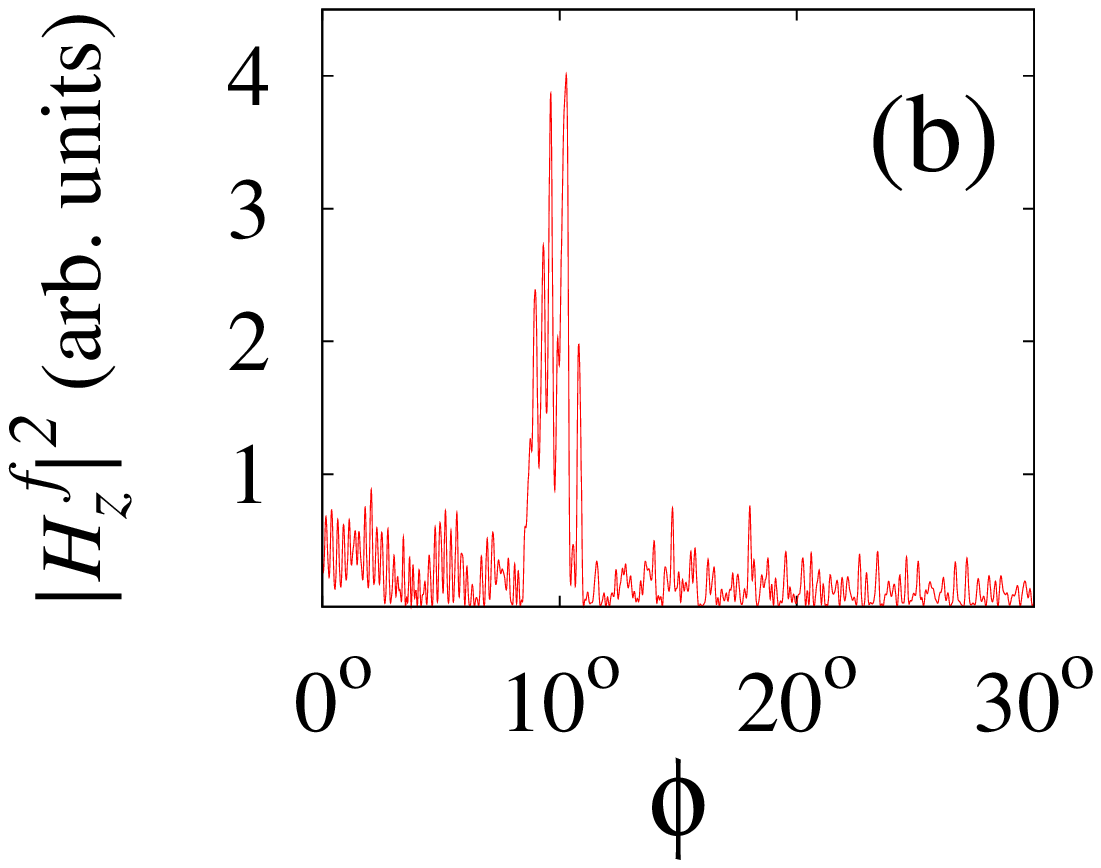}\\
  \end{tabular}
  \caption{\label{fig:fig3}  (Color online)
    (a) Far-field  radiation pattern $|H_z^f(\phi_j)|^2$ for the polymer 
    microcavity and high $Q$ period-4 MUPO mode shown in Fig.~\ref{fig:fig2}.
    (b) Magnification of $|H_z^f(\phi_j)|^2$ for 
    $0^{\circ} \le \phi \le 30^{\circ}.$
    The far-field pattern is strongly bidirectional.
    The angular resolution is $\Delta\phi = 0.008^{\circ}$.
  }
\end{figure}

Figure  \ref{fig:fig3} shows the far-field pattern generated by the
period-4 MUPO mode shown in Fig.~\ref{fig:fig2}.
The emission pattern is strongly bidirectional, with peaks at approximately 
$10^{\circ}$ and $170^{\circ}$.
The percentage $U$ of total emission within 
[8.5$^{\circ}$,11$^{\circ}$]$\cup$[169$^{\circ}$,171.5$^{\circ}$], is
$U \approx 25$\% which is $18$ times the corresponding value for isotropic emission.
From our prior discussion of internal reflection, much of the field 
leakage should be in the form of an evanescent field.
When the field hits the top of the hat, the evanescent field outside the
cavity is diffracted at various angles due to the curvature. 
A similar event takes place as the evanescent field along the
undercarriage diffracts at the corner of the hat. In both cases,  
there is no significant contribution to the primary peaks of the far-field 
pattern.

\begin{figure}
  \includegraphics[width=4.2cm]{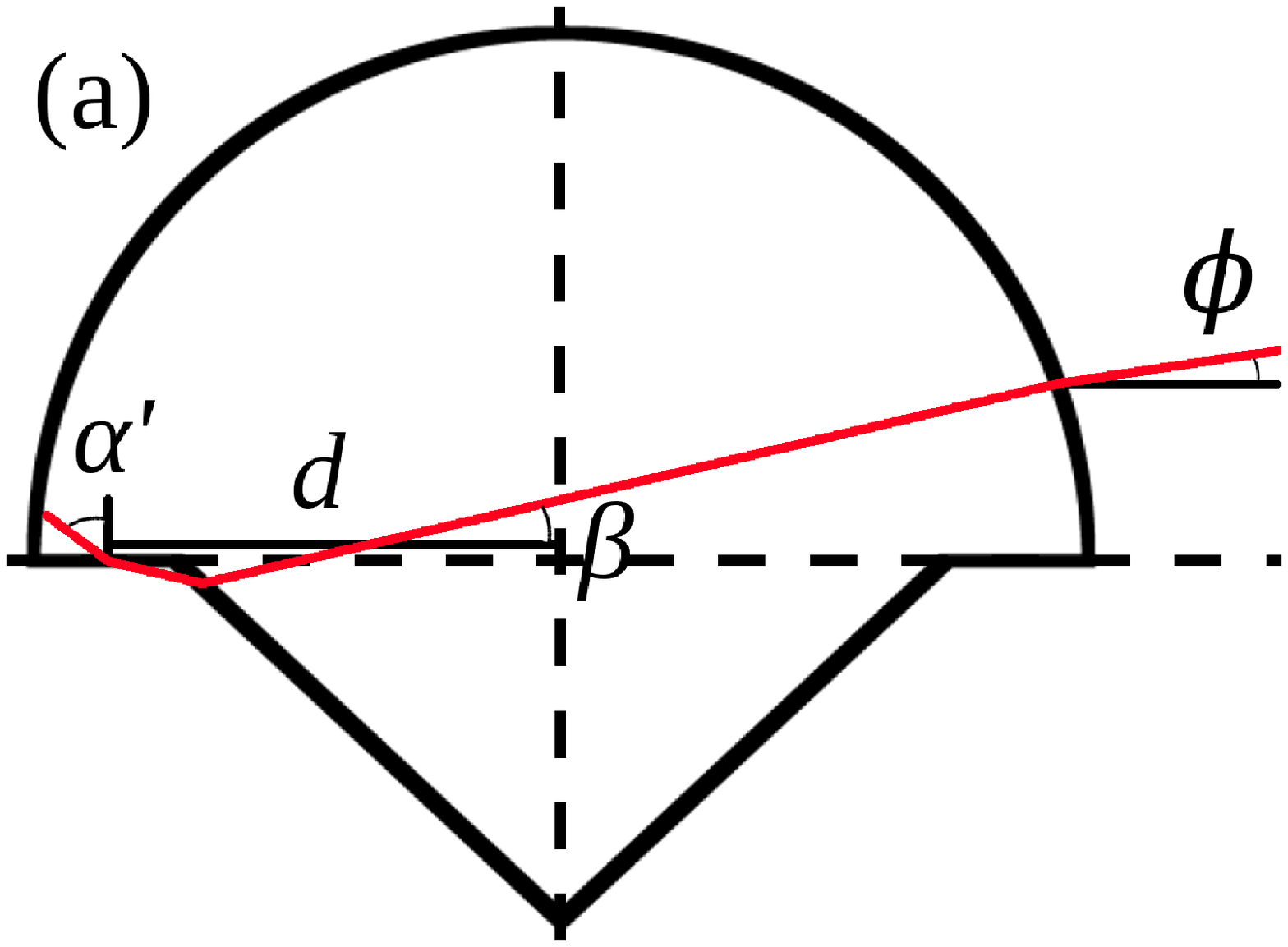} 
  \includegraphics[width=4.2cm]{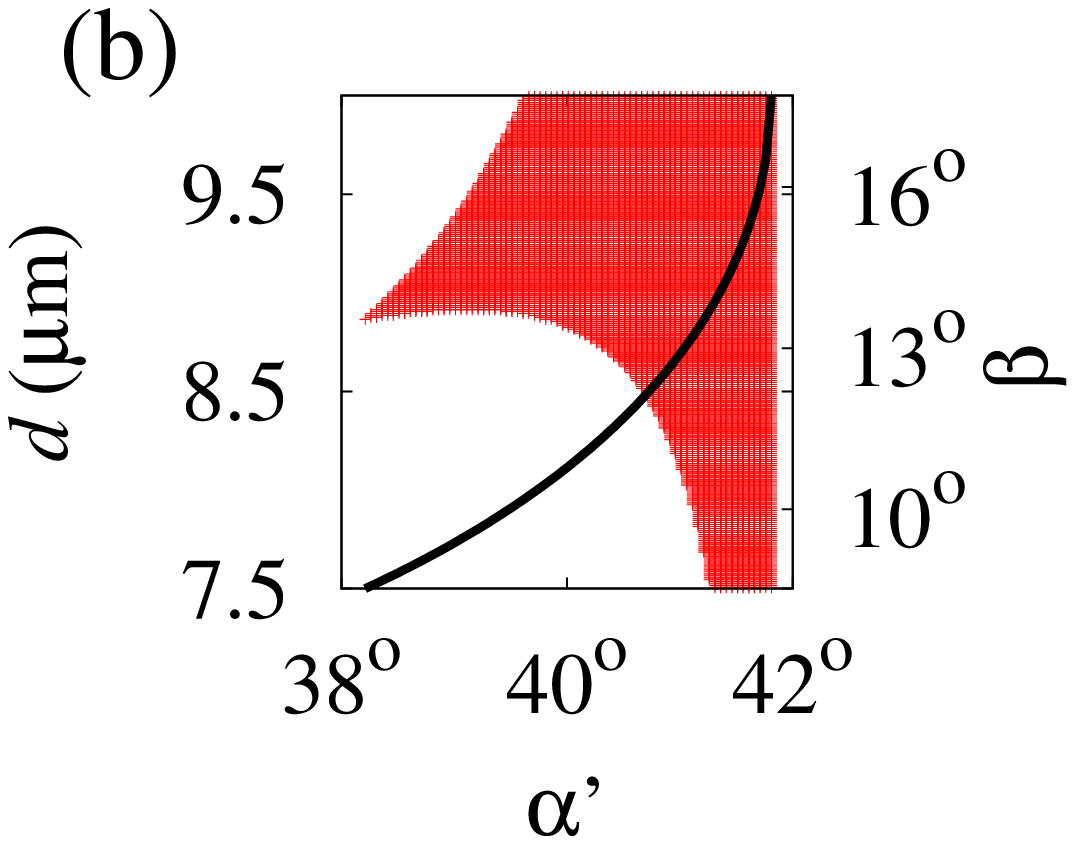}
  \caption{\label{fig:fig4} (Color online)
    (a) Calculated ray diagram indicating the point of incidence on the 
    undercarriage $d$, reentry angle $\beta$, and emission angle $\phi$ 
    in a mushroom cavity.
    (b) Values of $\alpha'$ and $d$ that result in emission angles 
    $8.5^{\circ} \le \phi \le 11^{\circ}$ (shaded area).
    The corresponding values of $\beta$ (black curve) 
    agree with the results shown in  Fig.\ \ref{fig:fig2}.
  }
\end{figure}

The directional emission is actually caused by refraction through
the undercarriage and subsequent penetration into the foot.
As suggested in Fig.~\ref{fig:fig2} and confirmed in Fig.~\ref{fig:fig4}, 
this is so because part of the MUPO's field strikes the boundary further 
up the curvature of the hat and hits the undercarriage with an incidence angle 
$\alpha' \lesssim \theta_c$.   
The field then refracts out of the cavity towards the foot.
Once the field enters the foot, it is refracted upwards to the boundary of
the hat and immediately escapes.
Figure \ref{fig:fig4}(a) shows the corresponding classical trajectory with 
$d$ marking the ray's incident location on the undercarriage, $\beta$ the 
upward angle of refraction once the ray reenters the cavity, 
and $\phi$ the emission angle.
Figure \ref{fig:fig4}(b) shows the values of $\alpha'$ and $d$ that traced 
along this trajectory satisfy $8.5^{\circ} \le \phi \le 11^{\circ}$.
Only incidence angles close to the critical angle---with $\alpha'=38^{\circ}$ 
being the minimum value found---result in emission angles agreeing with the 
far-field pattern shown in Fig.~\ref{fig:fig3}.
Moreover, the smallest incidence angles only result in agreement when 
$d > 8.8\mu$m, which is consistent with Fig.~\ref{fig:fig2} in that 
the mode is concentrated in the corner of the hat.
A measurement of the reentry angle from the Fig.~\ref{fig:fig2} data 
gives $\beta \approx 13.5^{\circ}$ (with some spreading),
which is also consistent with the calculation of $\beta$ in Fig.~\ref{fig:fig4}(b).

We now turn to semiconductor microcavities, which are computationally less 
demanding and less leaky and, as shown below, have high $Q$ modes in 
the optical range and can be used to further validate the mechanisms found 
in the polymer cavity.
With a refractive index  $n=3.3$, the semiconductor has a  smaller critical 
angle of $\theta_c \approx 17.6^{\circ}$.  
We consider this case for $R=10 \mu$m, $h=6.8 \mu$m, and $r=7.3 \mu$m, so the 
period-4 MUPOs are located just inside the chaotic region; we use a larger
grid spacing of  $\Delta x=8.3\,$nm and continue to focus on TE polarization 
(Table \ref{table1}).
One of the conditions for isolating the MUPO modes is no longer satisfied, 
namely $90^{\circ} - \theta^* \not< \theta_c$, rendering some of the regular 
orbits  to never reach the undercarriage with incidence angle smaller than 
$\theta_c$.
Nevertheless, because their incidence angles are farther apart from the critical 
angle by at least $\theta^*-\theta_p=1.9^{\circ}$, 
the period-4 MUPOs remain the highest $Q$ modes.

\begin{figure}[t]
  \begin{tabular}{rl}
    \multirow{2}{*}[1cm]{ 
      \includegraphics[width=3.5cm]{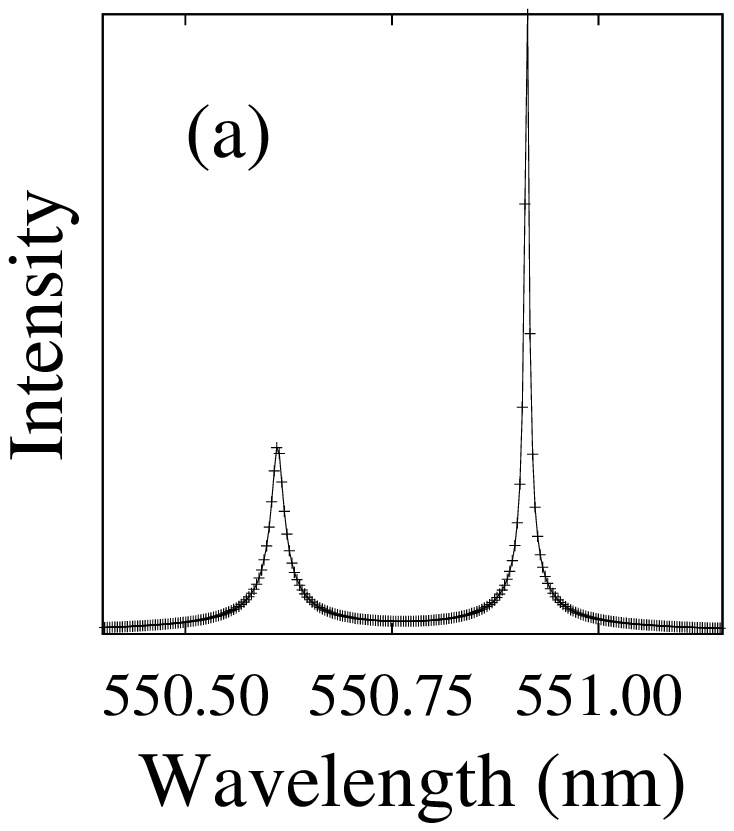}       
   } & \includegraphics[width=3.5cm]{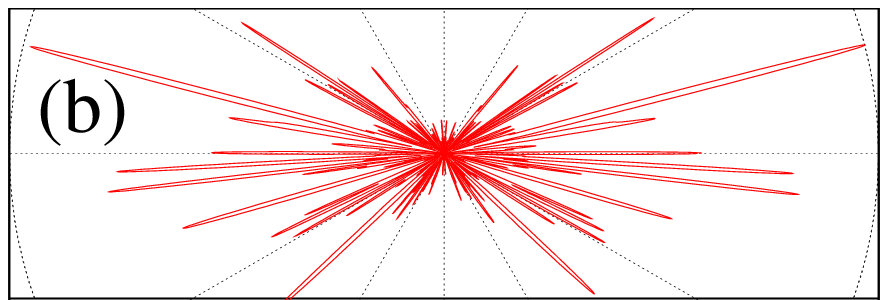} \\
      & \includegraphics[width=3.5cm]{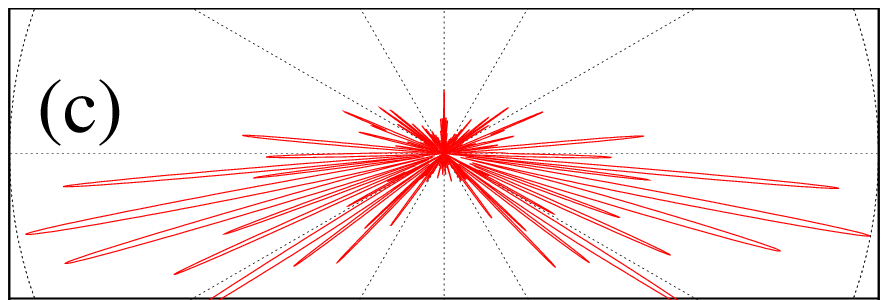} \\
  \end{tabular}
  \caption{\label{fig:fig5} (Color online)
    (a) Fourier transformation for the period-4 MUPO modes excited by 
    $\lambda = 550.92\pm 0.5\,$nm in a semiconductor microcavity. 
    There are two eigenmodes within this excitation range, at wavelengths 
    $\lambda = 550.61\,$nm and $\lambda = 550.91\,$nm.
    The $Q$ values for these modes are $39\,000$ and $140\,000$, respectively.
    (b,c) Far-field pattern $|H_z^f(\phi_j)|^2$ corresponding to 
    (b) $\lambda = 550.61\,$nm and (c) $\lambda = 550.91\,$nm. 
  }
\end{figure}

Indeed, with an excitation wavelength of $\lambda=580\pm 25\,$nm, the highest 
$Q$ eigenmodes are excited in the range $500\,$nm $< \lambda < $ $600\,$nm and
are all found to be period-4 MUPO modes irrespective of the foot shape. 
A harmonic inversion software \cite{mandelshtam_harmonic_1997} was used to 
estimate the $Q$ values. 
The highest $Q$ modes were selected and simulated individually 
for a more accurate estimate of $Q$.
A larger wavelength MUPO mode at $\lambda=598\,$nm has $Q=34\,000$ while
a smaller wavelength MUPO mode at $\lambda=526\,$nm has $Q=160\,000$.
Because resolution and simulation runtime are an issue for smaller wavelengths, 
the two eigenmodes identified in Fig.~\ref{fig:fig5}(a), at $\lambda=550.91\,$nm 
with $Q=140\,000$ and at $\lambda = 550.61\,$nm with $Q=39\,000$,
are chosen for further investigation.  
These two modes are representatives of the typical behavior found 
for the semiconductor cavity.
As shown in Figs.~\ref{fig:fig5}(b)-(c), the corresponding far-field pattern is 
close to isotropic for $\lambda = 550.61\,$nm and slightly bidirectional for 
$\lambda=550.91\,$nm.

\begin{figure}[t]
  \includegraphics[width=8cm]{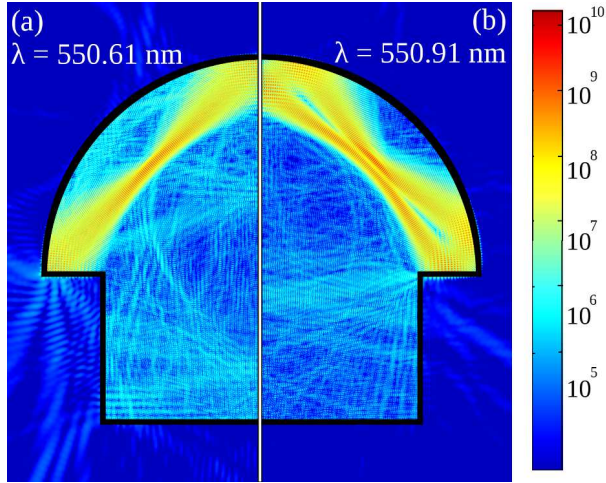} 
  \caption{\label{fig:fig6} (Color online) 
    (a,b) Spatial intensity $|H_z|^2$ for the period-$4$ MUPO modes
    at  (a) $\lambda = 550.61\,$nm, $Q = 39\,000$ and 
    (b) $\lambda = 550.91\,$nm, $Q = 140\,000$ in a semiconductor microcavity.
    The intensity was averaged over one oscillation period.
    Different times are selected to yield the same maximum intensity for 
    both plots.
  }
\end{figure}

Figure \ref{fig:fig6} shows the spatial intensity for these modes.  
The low $Q$ mode at $\lambda = 550.61\,$nm  is concentrated close to the corner
of the hat [Fig.~\ref{fig:fig6}(a)].
The evanescent field strongly diffracts at the corner resulting in a 
nearly isotropic far-field pattern. 
This case is thus similar to the one in Fig.~\ref{fig:fig2} and provides
further evidence that the directionality in the polymer microcavity originates 
from field reentry and subsequent refraction as opposed to direct refraction 
or diffraction of evanescent waves.
The high $Q$ mode at $\lambda=550.91\,$nm, on the other hand, is concentrated 
much closer to the intersection of the undercarriage and the foot, and hence 
has less leakage from diffraction at the corner [Fig.~\ref{fig:fig6}(b)].
The slight bidirectionality of the high $Q$ mode observed in Fig.~\ref{fig:fig5}(c) 
is caused by scattering of the field near the intersection.

The robustness of the high $Q$ period-4 MUPO mode shown in Fig. \ref{fig:fig6}(b) 
was tested by rounding the corners of the cavity with a radius of 
curvature $R_E$ and by introducing roughness to the radial boundary of 
the hat and undercarriage. 
The latter is defined by a standard deviation $A$ and modulation period $D$.
As the perturbation increases, either through rounded corners or roughness 
in the boundary, Fourier transformations reveal that the wavelength of the 
MUPO mode changes little.
The largest roughness for which the MUPO mode is not destroyed is for $A=9\,$nm 
and $D=40\lambda$.
Even in this case, the resulting wavelength shift was just  0.01\%.
Similar or even more pronounced robustness was observed for the directionality 
of the far-field pattern of the highest $Q$ mode of the polymer 
microcavity \cite{comment}. 
However, the $Q$ values proved quite sensitive to the perturbations.  
For the case Fig. \ref{fig:fig6}(b), the $Q$ value reduces from $140\,000$ to 
$73\,000$ for $R_E$ = $300\,$nm and to $Q=3\,200$ for  $A=9\,$nm and $D=40\lambda$.
Nevertheless, the spatial intensity pattern of the MUPO mode is fairly well 
preserved in all cases considered.
Indeed, unlike the classical case, where a MUPO either exists or does not exist, 
the semiclassical MUPO mode changes with the perturbations but mimics the 
unperturbed MUPOs for the entire lifetime of the mode in the cavity. 

MUPOs are inherently less dispersive than unstable periodic orbits and, 
as shown here, underlie modes that can be isolated and trapped in open optical cavities.
The openness and non-convexity of the cavity, and consequent reentry of 
refracted electromagnetic fields, can lead to directional emission,
fundamentally altering the far-field radiation pattern.
The recently demonstrated prevalence of MUPOs in many billiards \cite{altmann08} 
indicates that these findings are most likely not limited to the mushroom 
billiards considered in this Letter.

The authors thank M.~Sukharev for stimulating discussions. 
This work was funded by 
NIST Grant 70NANB6H6162, NSF Grant DMR-0808937, 
NSF-MRSEC program DMR-0520513, and
supported by the Yale Biomedical HPC Center
and NIH Grant RR19895.


\begin{thebibliography}{}
\bibitem{comment}
To lighten the computational burden, these comparisons
were run using a 
coarser grid with $\Delta x = 5\,$nm.
Though numerical dispersion may influence the calculations, the qualitative 
character of the results remains the same.
\end{thebibliography}
\end{document}